# WEB SITE OPTIMIZATION THROUGH MINING USER NAVIGATIONAL PATTERNS

**Biswajit Biswal**

Oracle Corporation
biswajit.biswal@oracle.com

***ABSTRACT***

*With the World Wide Web (www)'s ubiquity increase and the rapid development of various online businesses, the complexity of web sites grow. The analysis of web user's navigational pattern within a web site can provide useful information for server performance enhancements, restructuring a website and direct marketing in e-commerce etc. In this paper, an algorithm is proposed for mining such navigation patterns. The key insight is that users access information of interest and follow a certain path while navigating a web site. If they don't find it, they would backtrack and choose among the alternate paths till they reach the destination. The point they backtrack is the Intermediate Reference Location. Identifying such Intermediate locations and destinations out of the pattern will be the main endeavour in the rest of this report.*

## 1. INTRODUCTION

From a user's perspective, hypertext links on the web page form a directed graph between distinct information sources. A website is a collection of web pages forming a hierarchically nested graph (see Figure. 1). A web site generally has a "root page" from which there should be author-designed paths to all local content. However different users have different needs. The same user may need different information at different times. A web site may be designed in a particular way, but be used in many different ways. Therefore, it is hard to organize a web site such that pages are located where users expect to find them.

In this paper, an algorithm is proposed to identify all the destination pages in a web site whose location is different from the location where users expect to find them. The key insight is that users will backtrack if they do not find the page where they expect it. The point from where they backtrack is the Intermediate Reference Location (IRL) for the page. IRL's with maximum hits will then be made to include navigation links to the destination page. It is also worth mentioning that users may try multiple IRL for a destination page.

User navigational patterns can be studied from the web access-logs generated by the system. Web access-logs record the access history of users that visit a web server. Web servers register a web log entry for every single access they get, in which important pieces of information about accessing are recorded, including the URL requested, the IP address from which the request originated, and a timestamp. A sequential access-pattern is generated out of these logs. A sequential access-pattern represents an ordered group of pages visited by users. Mining of these access-patterns will lead to the identification of user' behaviour and thus the solution.

The rest of this paper is organized as follows: Section 2 gives an overview of the related work, whereas the mining of sequential access-patterns is described in Section 3. Section 4 contains the experimental results, and finally Section 5 concludes the paper.

## 2. RELATED WORK

A lot of work has been done in web access pattern mining.

Pei et al. [5] propose Web access pattern tree or WAP-tree for efficient mining of access patterns from web logs. The WAP-tree can store highly compressed, critical information and facilitates the development of novel algorithms for mining access patterns in large set of logs.

Perkowitz et al. [3 & 4] present adaptive web sites, which is the semi-automatic creation of web pages. Their PageGather algorithm finds pages that co-occur frequently and create an index page which consists of all the links to these pages.

Nakayama et al. [6] discover the gap between web site designer's expectations and user's behaviour by measuring the inter-page conceptual relevance and access co-occurrence. The technique suggests a multiple traversal frequently from page layout features.

Shahabi et al. [2] propose a profiler, which captures client's selected links and pages order, accurate page viewing time and cache references, using a Java based remote agent, which is then utilized by a knowledge discovery technique to cluster users with similar interests.

Shiliopoulou et al. [8] compare navigation patterns of registered users with those of unregistered users. This comparison leads to rules on how the site's topology should be improved to turn non-customers into customers.

However one drawback of these approaches is that they work on some dependencies with the client side browser or the other. Some of these approaches suggest no clues to improve the web site structure and concentrate less on the user's behaviour.

## 3. MINING WEB ACCESS-LOGS

As mentioned earlier, web pages are linked together and users travel through them back and forth in accordance with the links and icons provided. Therefore, some node might be visited only because of its location, not content. Consequently, such backwards traversals should be taken into consideration in the research to study user's behaviour.

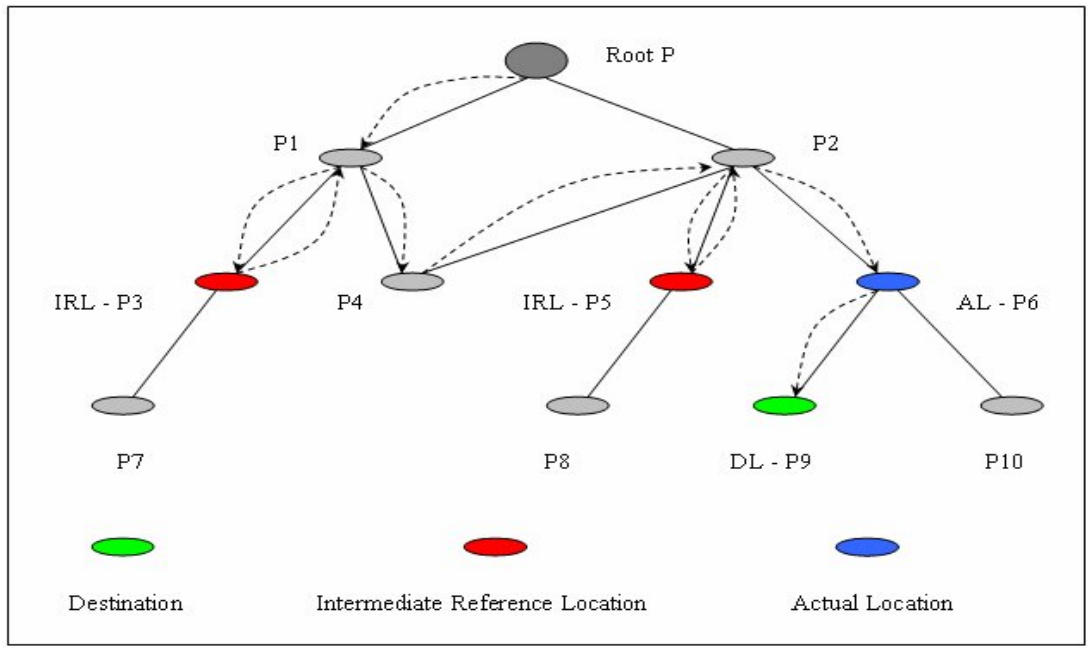

**Figure 1: Web Site Structure & Navigational Pattern**

## 3.1. User Navigational Patterns

Figure 1 illustrates the web site structure and the user navigational pattern. The nodes in the traversal graph indicate the web pages that are visited in a browsing session. The user starts from the root node P. There are two links for pages P1 & P2 in the root. The user proceeds with P1 due to some probability of his/her own. From P1 the user navigates to P3 (ignore P4). However P7 is not the intended page. So the user backtracks and goes to P1 to P4. Then the user moves to P2. P2 has two links for pages P5 & P6. The user opts for P5 and again backtracks. Finally the user moves to P6 and then the destination page P9. The pattern is as follows:

$$P \rightarrow P1 \rightarrow P3 \rightarrow P1 \rightarrow P4 \rightarrow P2 \rightarrow P5 \rightarrow P2 \rightarrow P6 \rightarrow P9$$

## 3.2. Identifying Destination and Intermediate Reference Locations

**Single Destination Page:** Here the user is looking for a single destination page. It starts from the root node. The user chooses the link that appears most likely to lead to Destination. If any of the page is not the Destination, the user will backtrack and go to some other page that has maximum probability.

*While (not empty (Page list) || current location p is not the destination page P)*

*Choose a link from p that seems most likely to lead to P.*
*Backtrack to the parent of p with some probability.*

**Multiple Destination Pages:** Consider the case where the user has a set of Destination pages $P_1\ldots P_n$. The pattern is similar, except that after finding (or giving upon) $P_i$, the user then starts looking for $P_{i+1}$.

*While (not empty (Page list) || current location P is not the destination page $P_i$*

*Choose a link from P that seems most likely to lead to $P_i$*
*Backtrack to the parent of P with some probability or give up on $P_i$ and look for $P_{i+1}$*

The pattern illustrated in Figure: 1 might be of Single Destination or multiple Destinations. It is hard to identify by looking at the web log. The only distinguishing factor would be the time spent at each of these locations. The average time spent at each page by all the users can then be evaluated and become a delimiting factor for this page to be an Intermediate Reference Location or a Destination Location. However, it is likely that a starter (who has spent less time on the site) might spent more time in the page than the time spent by an veteran (who has spent more time on the site) user. So, the total time the user has spent in the web site prior to coming to this page should also be considered. The threshold time for each page could be evaluated as follows:

*Let assume that Page P has $t_1, t_2\ldots t_n$ times spent by n users. These users have spent $T_1, T_2,\ldots T_n$ times on the web site before coming to this page respectively. The parameter "đ" is a damping factor which can be set between 0.15 and 0.85. Its value is proportional to the popularity of the page in the web site which can be set by the web site administrator. The parameter is valued so to better approximate the time stated by the log.*

$$T_P = đ\ (t_1 T_1 + t_2 T_2 + \ldots + t_n T_n)\ /\ (T_1 + T_2 + \ldots + T_n)$$

## 3.3. Sources

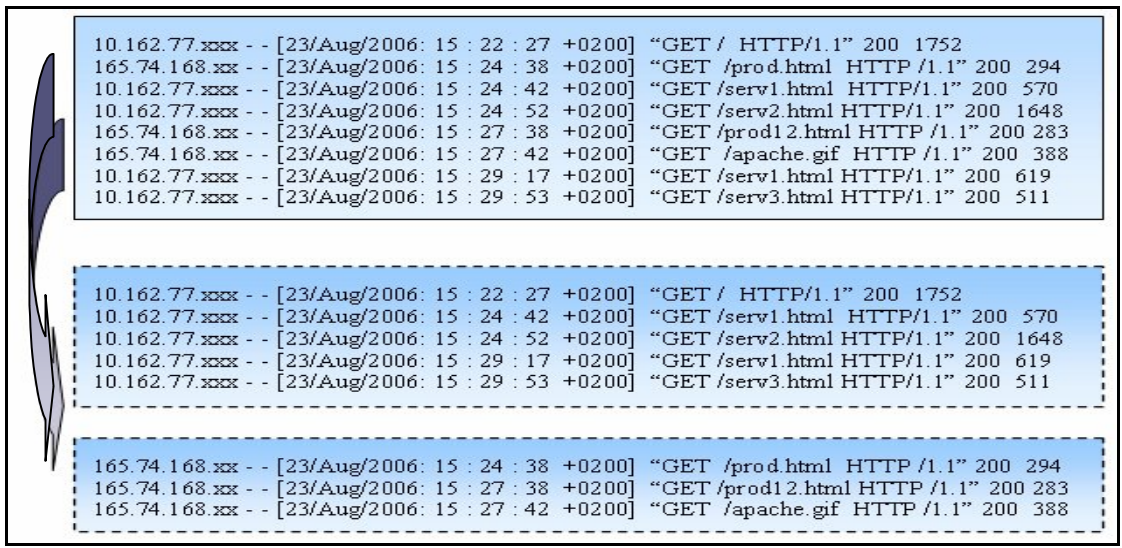

**Figure 2: Example of a Web Access Log**

Let's look at a typical web access log in Figure 2.

Each web log entry represents each user's access to a web page and contains the user's IP address, the Timestamp, the URL address of the requested object, and some additional information. Access requests issued by a user within a single session with a web server constitute a user's access sequence. These data sets commonly used for web traversal mining are collected at the server-level, proxy-level or client-level. Each data source differs in terms of format, accuracy, scope and method of implementation.

Client-level logs hold the most accurate account of user behaviour over www. If a client connection is through an Internet Service Provider (ISP) or is located behind a firewall, its activities may be logged at this level. The primary function of proxy servers or firewalls is to serve both as a measure of security to block unwanted users or as a cache resource to reduce network traffic by reusing their most recently fetched files. Their log files may include many clients accessing many Web Servers. In the log files, their client request records are interleaved in their received order. The process of logging is automatic and requires less intervention.

The web access log can be specialized to different sets of patterns based upon the IP address and Time stamp as shown in Figure 2. The last two blocks consists of entry for a single client sorted by the timestamp. These extracted patterns can then be indexed to a database or to some temporary buffer for mining. Note that only the html pages are considered for the research work. So, all the other objects (jpg, gif, etc.) accessed by the users are ignored from the pattern.

### *3.4. Algorithm*

Let, Array page_array contains the ordered set of pages {P1, P2, P3 …. Pn}.
Function Mark (Page, Indicator) marks the page as specified in the indicator parameter (DL or IRL).     Function IsConnected (Page1, Page2) returns True if Page1 has a link to Page2 or False otherwise.

*For i: = 1 to n-1 begin*

  *if( ( page_array [i-1] ) = = ( page_array [i+1] )  ||  ( IsConnected (page_array[i] ,*
                  *page_array [i+1] ) ) )*
  *if( t ( page_array [i] ) ≥ T (page_array [i] ) ) then Mark ( page_array [i] , DL );*
                  *else Mark ( page_array [I] , IRL ) ;*

*end;*

Now the pattern will have a list of IRL's followed by a DL, and so on. The Actual Location (AL) is the page traversed before the DL.

*For each DL in page_array begin   Add (DL, IRL List , AL) to the table.*

*End;*

The algorithm checks for the previous and next page. If the pages are same, then the current page could either be an IRL or a DL. It also checks for the absence of a link from the current to the next page. In that case the user might have used a navigation link or hit the back button to go to the next page. In either of the case the page is either an IRL or a DL. Next, the algorithm compares the time currently spent at the page with the threshold time. If the current time spent is greater than the threshold, then the page is a Destination Location else an Intermediate Reference Location.

## *3.5. Optimization*

From the above algorithm a table is created with the following columns: D, AL, $Ĩ_1$, $Ĩ_2$… where D is the Destination Location, AL is the Actual Location and $Ĩ_1$, $Ĩ_2$…are the list of IRL's. The table structure is not normalized and can be ignored as it's out of scope of this report.

| D  | AL | $Ĩ_1$ | $Ĩ_2$ | $Ĩ_3$ | $Ĩ_4$ |
|----|----|----|----|----|----|
| D1 | A1 | P3 | P5 | P4 | P1 |
| D1 | A1 | P2 | P1 |    |    |
| D1 | A1 | P1 |    |    |    |
| D1 | A1 | P3 | P4 | P2 |    |

Let,

$β_k$, be the probability of finding the destination in page Pk.
$Ω_n$, be the probability of finding the destination in nth IRL. $Ω$ values can be set as 1, 0.75, 0.5, 0.25 for $Ĩ_1$, $Ĩ_2$, $Ĩ_3$, $Ĩ_4$ respectively.
$S_P$ represent the average probability of finding the destination for any IRL.
P1, P2, P3 … Pk represents k pages. $Ĩ_1$, $Ĩ_2$, $Ĩ_3$, $Ĩ_4$ ….. $Ĩ_n$ represents n IRL's.

$$β_k := \sum Ω \text{ (Pk position in each record)} , \quad S_P := \sum β_k / \sum k$$

*For each Record in Table begin*
*For i := 1 to n begin    Increment β value for each IRL as per $Ω_n$   end*
*end*

$S_P = \sum β_k / \sum k$ ;
$P = Max(β_k)$;

*For each page in IRL_List  begin if $β_k ≥ S_p$ then  Mark the page Pk as recommended. end*

*end;*

*For each Record in Table begin*
*For i := 1 to n begin    If $β_k ≥ S_P$ then Remove $Ĩ_k$, $Ĩ_{k+1}$,… $Ĩ_n$ from record. end*
*end*

The IRL with the maximum probability is marked as recommended page and all the subsequent pages are removed from the table. The table mentioned above is evaluated as follows:

$\beta_1 = 0.25\ (1^{st}\ \text{Row}) + 0.75\ (2^{nd}\ \text{Row}) + 1\ (3^{rd}\ \text{Row}) = 2$

$\beta_2 = 1\ (2^{nd}\ \text{Row}) + 0.25\ (4^{th}\ \text{Row}) = 1.25$

$\beta_3 = 1\ (1^{st}\ \text{Row}) + 1\ (4^{th}\ \text{Row}) = 2$

$\beta_4 = 0.5\ (1^{st}\ \text{Row}) + 0.75\ (4^{th}\ \text{Row}) = 1.25$

$\beta_5 = 0.75\ (1^{st}\ \text{Row}) = 0.75$

$S_P = (2+1.25+2+1.25+0.75)/5 = 1.45$

P1 & P3 pages are recommended and the new table is constructed below:

| D | AL | $Ĩ_1$ | $Ĩ_2$ | $Ĩ_3$ | $Ĩ_4$ |
|---|----|----|----|----|----|
| D1 | A1 | | | | |
| D1 | A1 | | P2 | | |
| D1 | A1 | | | | |
| D1 | A1 | | | | |

The algorithm identifies the IRL that has maximum probability of attempt for any user. This IRL can then be made to include navigation links to the destination page. The recommended IRL now becomes one of the Actual Location for the Destination page. Other way is to restructure the web site using a similarity matrix on these extracted pages.

## 4. EXPERIMENTAL RESULTS

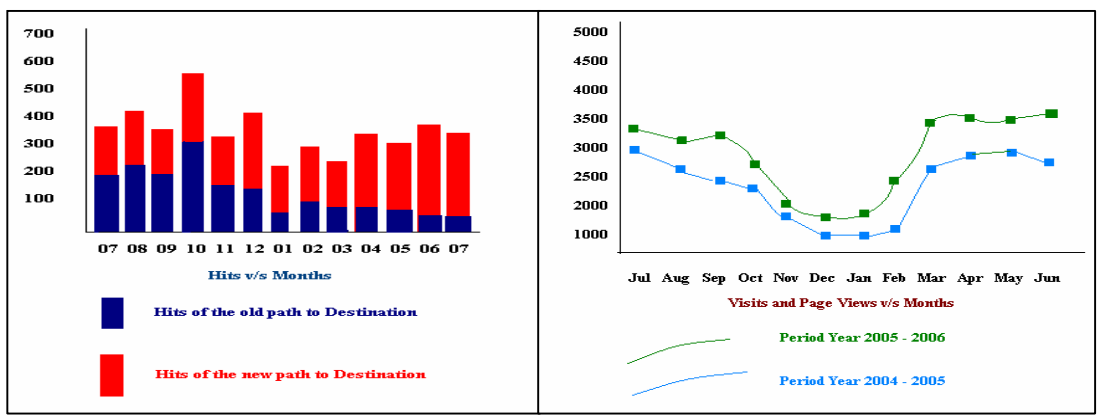
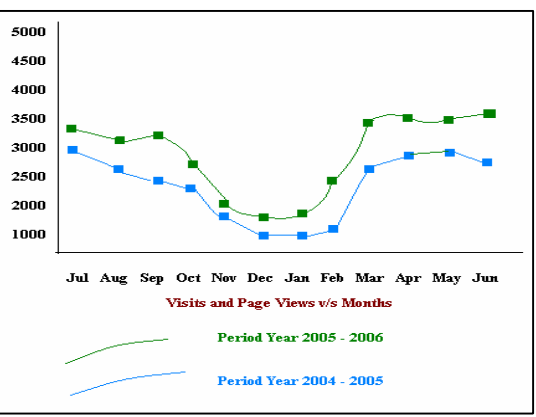

**Figure. 3**  **Figure. 4**

The experiments were carried out for a period of one year on a real online ordering system of a major telecom corporation of US. The web site was restructured and optimized after analysing the web access logs.

Figure. 3 shows the number of hits to that destination page for every month from period Jul-2005 till Jul – 2006. The blue (dark) block illustrates the number of hit to the destination page through the old path before optimization and the red (light) block illustrates the number of hits to the destination page through the additional new path after optimization. It clearly suggests that more users are visiting the new path.

Figure. 4 show the total number of visits for a period of 2 years. The blue (lower) line shows the variance of user visits from period Jul 2004 – Jul 2005 and the green (upper) line from period Jul 2005 – Jul 2006. The users are authenticated customers (not just visitors) who used the website to process their orders. It can be inferred from the graph that total number of users has increased after the web site optimization. It is calculated that the percentage of customers, who use the online system for placing order, has increased during the period.

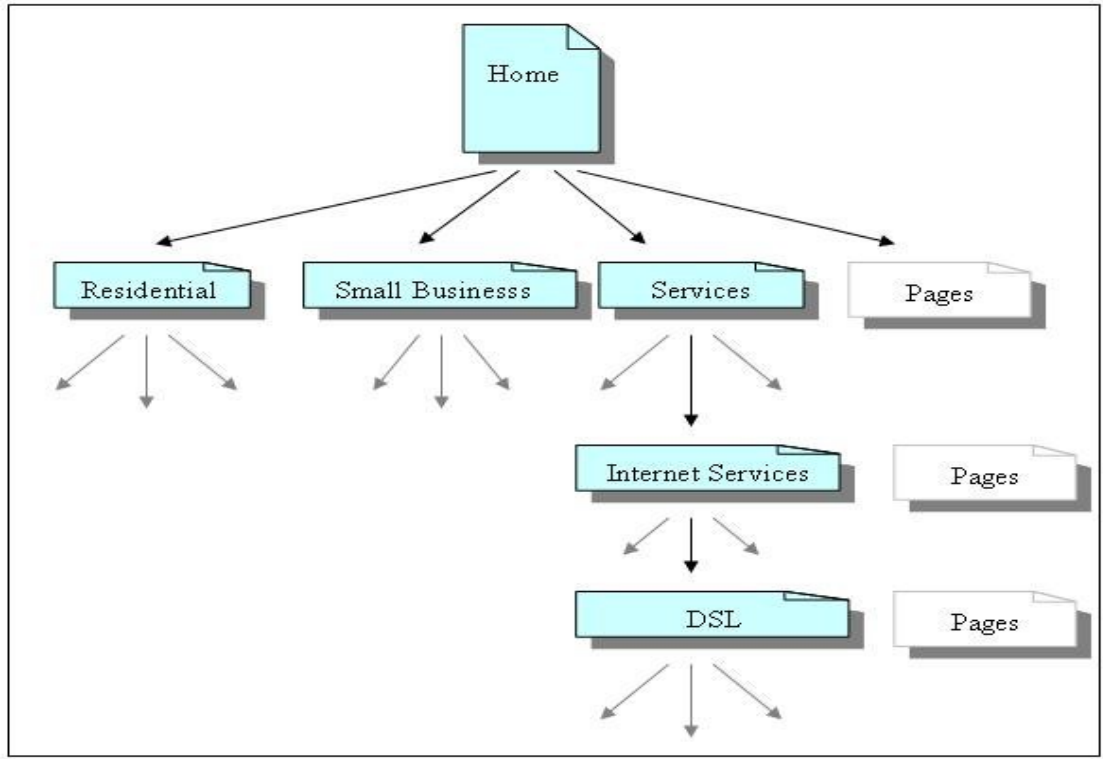

**Figure 5: Website Structure before optimization**

Figure 5 shows the earlier website structure before optimization. The research was focus around this level deep of pages and the pattern was gathered till this level. Users who process their orders at the service pages are considered for this research. The service pages at Level 4 were considered as the leaf pages and thus the Destination Location. All other pages other than the root page can be a Destination Location as per the analysis.

A single observation is mentioned below:

*Destination Location: Internet Services Page*
*Intermediate Reference Location:  Residential & Small Business page*
*Actual Location: Services page*

The user expects to find the "Internet Services" page in the "Residential" page or "Small Business" page instead of "Services" page. Similarly, in other observations it is noticed that users enters the "residential" or "small business" page and expects to find all the services offered under that group. According to the experimental results, around 20% of the destination pages have Intermediate Reference Locations different from their Actual Locations. On an average each service page has thousands of visitors among which potential users are in hundreds.

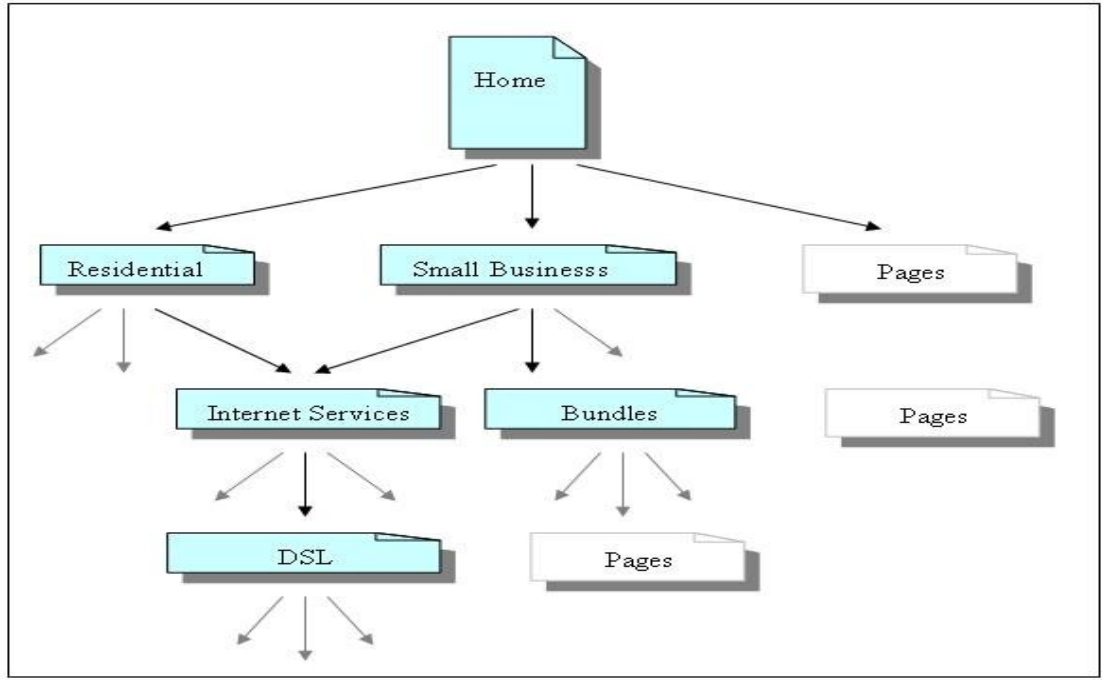

**Figure 6: Website Structure after optimization**

Figure 6 shows the new website structure after optimization. The various available services like: internet services & bundle are put under the respective Residential & Small Business pages and the Service section is removed from the structure.

## 5. CONCLUSIONS

In this study, an algorithm is proposed for mining user navigational patterns through web access-logs to the advantage of web site owner. The Intermediate Reference Locations and the destinations are identified taking into account user identification, page viewing time, web site viewing time, etc. The performance of the proposed algorithm is examined experimentally with real and synthetic data.

As a future work, it will be interesting to explore if there are better approaches to identify IRL and DL accurately. One suggested approach would be to analyse the content of web pages to find out similarities. Finally, predictive analytics model can be used to better forecast specific user action/behaviour from access-patterns.